\documentstyle[aps,prl,multicol,graphicx]{revtex} 

\begin{document}

\title{Doubling of the bands in overdoped 
$Bi_{2}Sr_{2}CaCu_{2}O_{8+\delta}$-probable evidence for c-axis 
bilayer coupling}

\author{Y. -D. Chuang$^{1,2}$, A. D. Gromko$^{1}$, A. 
Fedorov$^{1,2}$, D. S. 
Dessau$^{1}$, Y. Aiura$^{3}$, K. Oka$^{3}$, Yoichi Ando$^{4}$, H. 
Eisaki$^{5}$, S.I. Uchida$^{5}$} 
\address{$^1$Department of 
Physics, University of Colorado, Boulder,Colorado, 80309-0390} 
\address{$^2$Advanced Light Source (ALS), Lawrence Berkeley National 
Laboratory (LBNL), Berkeley, CA 94720}
\address{$^{3}$Electrotechnical Laboratory (ETL), 1-1-4 Umezono, 
Tsukuba, Ibaraki 305-8568, Japan}
\address{$^{4}$Central Research Institute of Electric Power Industry 
(CRIEPI), 2-11-1 Iwato-Kita, Komae, Tokyo 201-8511, Japan}
\address{$^{5}$Department of Superconductivity, University of Tokyo, 
7-3-1 Hongo, Bunkyo-ku, Tokyo 113, Japan}

\maketitle


\begin{abstract}
We present high resolution ARPES data of the bilayer superconductor 
$Bi_{2}Sr_{2}CaCu_{2}O_{8+\delta}$ (Bi2212) showing a clear doubling of the 
near $E_{F}$ bands. This splitting approaches zero along the 
$(0,0) \rightarrow (\pi,\pi)$ nodal line and is not observed in single layer 
$Bi_{2}Sr_{2}CuO_{6+\delta}$ (Bi2201), suggesting that the splitting 
is due to the long sought after bilayer splitting effect. The 
splitting has a magnitude of approximately $75meV$ near the middle of 
the zone, extrapolating to about $100meV$ near the $(\pi,0)$ point. 

\end{abstract}
\vspace*{-0.3 in}
\begin{multicols}{2}
\narrowtext 

One of the central features of the cuprate superconductors is that 
their physical properties depend strongly on the number of $Cu-O$ 
planes (n) per unit cell.  For example, increasing n from 1 to 3 
almost universally results in large enhancements in the 
superconducting $T_{c}$ \cite{Tarascon}.  This would seem to argue for the 
importance of some sort of coupling between the $Cu-O$ planes within a 
unit cell.  Despite this, there have been essentially no direct 
observations of the intracell coupling between the planes, and the 
vast majority of theoretical models for the cuprates focus on a single 
$Cu-O$ plane, i.e.  on the limit of zero intra and intercell c-axis 
coupling.

In this letter, we present the first high energy/momentum resolution 
angle resolved photoemission (ARPES) data of a cuprate superconductor 
which shows a clear doubling of the bands in the normal state.  The 
splitting is observed in the double layer compound 
$Bi_{2}Sr_{2}CaCu_{2}O_{8+\delta}$ (Bi2212) but is much weaker or 
absent in the single layer compound $Bi_{2}Sr_{2}CuO_{6+\delta}$ 
(Bi2201).  The splitting approaches zero along the 
$(0,0)\rightarrow(\pi,\pi)$ nodal line and is maximal at 
$\bar{M}(\pi,0)$, and appears to be larger for overdoped samples.  
These trends point towards an intracell c-axis coupling between the 
$Cu-O$ planes (bilayer splitting) as the likely origin of the 
splitting, although other effects such as microscopic phase separation 
may play a role.  In addition to the new physics of this coupling, our 
new findings may help resolve the recent controversy about the 
electron-like \cite{Chuang,e-DongLai,Pasha} and hole-like 
\cite{Ding,Borisenko,Fretwell} Fermi surfaces which have recently been 
measured on Bi2212.

The experiments were carried out at the Stanford Synchrotron Radiation 
Laboratory (SSRL), Stanford, CA, at the Advanced Light Source (ALS), 
Berkeley, and at the Synchrotron Radiation Center (SRC), Stoughton, 
WI. At each facility we used a Scienta 200mm energy analyzer running 
angle mode to simultaneously collect up to 84 individual Energy 
Distribution Curves (EDCs) along an $\approx14^{o}$ angular slice.  The angle 
resolution was about $0.08^{o}$ along the angular slice (the $\theta$ 
direction) and about $0.4^{o}$ in the perpendicular direction (the 
$\phi$ direction).  At the experimental photon energy $h\nu=24.7eV$, 
the Brillouin Zone edge $(\pi/a)$ will be $\approx 20^{o}$ and the 
corresponding momentum resolution will be $(\Delta k_{x}, \Delta k_{y}) \approx 
(0.008\pi/a, 0.04\pi/a)$ with $a$ the lattice constant.  The energy 
resolution was better than $20meV$ determined from the $10-90\%$ width 
of a gold reference which was in electric contact with the sample and 
was used to determine the Fermi Energy $E_{F}$.  Data is shown from a 
lightly overdoped Bi2212 sample with a $T_{c}=85K$ and a lightly 
overdoped Bi2201 samples with $T_{c} \approx 5K$.  The photon 
polarization direction was along the $\theta$ direction and the 
$\Gamma(0, 0)- Y(\pi,\pi)$ high symmetry line.  All data shown here 
was normalized by using the spectral weight in the high binding energy 
window $(-0.8eV, -0.6eV)$ only \cite{normalization}.

\begin{figure}[htbp]
  \begin{center}
    \includegraphics[scale=0.55]{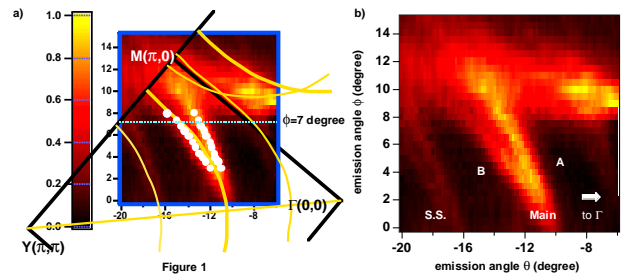}
    \caption{
(a),(b) ARPES intensity map within $\pm 5 meV$ of the 
  $E_{F}$ from the normal state $(T=100K)$ of an overdoped 
  Bi2212 sample.  The photon energy was $24.7 eV$ and the polarization 
  $40\%$
  along the $\theta$ direction and $60\%$ out of plane.  An overlay of 
  the data with the Bi2212 Brillouin Zone is shown in (a).  The 
  hole-like main FS (labeled "Main") plus superstructure bands 
  (labeled "S.S.") was obtained at $19 eV$ by Ding {\it et al.} [5] 
  (yellow lines).  The FS crossings determined from MDC fitting are 
  indicated by white dots.}
 \end{center}
\end{figure}

Figure 1$\bf a$ shows the schematic plot of the commonly accepted 
Fermi Surface (FS) topology of Bi2212 \cite {Ding,Borisenko,Fretwell} 
in part of the first Brillouin Zone.  The thick yellow lines are the 
main FS and the thin yellow lines are the FS produced by the crystal 
distortion (the superstructure, labeled ``S.S.'').  Figure 1$\bf b$ is 
the intensity distribution of the ARPES spectral weight at $E_{F}$ in 
the normal state of a lightly overdoped Bi2212 sample, with the $\vec 
k$ space location of this panel indicated in figure 1$\bf a$.  Modulo 
matrix element effects, the high intensity locus at $E_{F}$ should 
represent the FS topology \cite{Aebi}.  In panel $\bf b$, one can see 
that the main FS splits into two branches, one produced by band A and 
the other by B. This splitting behavior can also be seen in the 
superstructure FS, even though the intensity is much weaker there.  
This behavior is not expected within the generic FS topology shown by 
the yellow lines in figure 1$\bf a$.

In order to examine the splitting of these bands more carefully, we 
examine the $\vec k$-dependence of the states as a function of binding 
energy.  Figure 2$\bf a$ shows data along the blue line $(\phi=7^{o})$ 
in figure 1$\bf a$, with the main band crossing $E_{F}$ near the 
center of the window.  The so-called "main band", which shows up as 
the high intensity feature, disperses from binding energy $\approx 
-0.12eV$ (around emission angle $\theta=-6^{o}$) towards $E_{F}$.  
Around $\theta=-12^{o}$, another weaker feature appears at slightly 
deeper binding energy.  This two-peak structure (labeled ``A'' and 
``B'' here) can be easily seen in the energy distribution curve (EDC) 
shown in panel $\bf b$, which was taken at $\theta=-12^{o}$ (vertical 
white line in panel $\bf a$).  We emphasize that such two-peak structure 
in EDC is self-evident without any analysis on the data.  However, further 
quantitative determination of the peak positions, FS crossings, or 
splitting is difficult from the EDCs, as a detailed understanding of 
their line shape and background is still unclear.  In addition, the 
Fermi function distorts the line shape near the Fermi energy.

On the other hand, it is now well established that for relatively 
dispersive states the Momentum Distribution Curves (MDCs, intensity 
versus emission angle $\theta$ with fixed binding energy) have a 
simple line shape which can be modeled by a simple Gaussian or 
Lorentzian function \cite{Valla,Adam,Kaminski}.  This is equivalent to 
saying that the electron self-energy is essentially $\vec 
k$-independent over the narrow $\vec k$-range of a MDC peak.  A 
further advantage of the MDC method is that it won't be affected by 
the Fermi function cutoff, so in principle one can track ARPES 
features all the way up to or even above $E_{F}$.  Figure 2$\bf c$ 
shows the MDCs from the data of 2$\bf a$ over the full $\theta$ range 
and from $10 meV$ above $E_{F}$ to $60 meV$ below.  The MDC at $E_{F}$ 
is further shown in panel $\bf d$ (blue dots), along with a fitting 
(red line) of the main band into two separate Lorentzian peaks A and B 
and the superstructure band into two corresponding features (green 
lines), plus a linear background (black dash line).  For the fit, the 
angular width, position, and intensity of each feature was allowed to 
vary.  It was found that the angular width of both branches were 
essentially identical, and the intensity ratio and the angular 
splitting between the branches found to be roughly constant as a 
function of binding energy (see figure 2$\bf e$).  The high quality 
fits allow an accurate determination of the peak positions for each 
component.  Such fits were performed for each MDC at $E_{F}$ between 
the $\phi$ angles of $3^{o}$ and $8^{o}$ \cite{range}.  In this way 
the white dots of figure 1$\bf a$ were determined which, as expected, 
perfectly match the high intensity locus.

\begin{figure}[htbp]
  \begin{center}
    \includegraphics[scale=0.8]{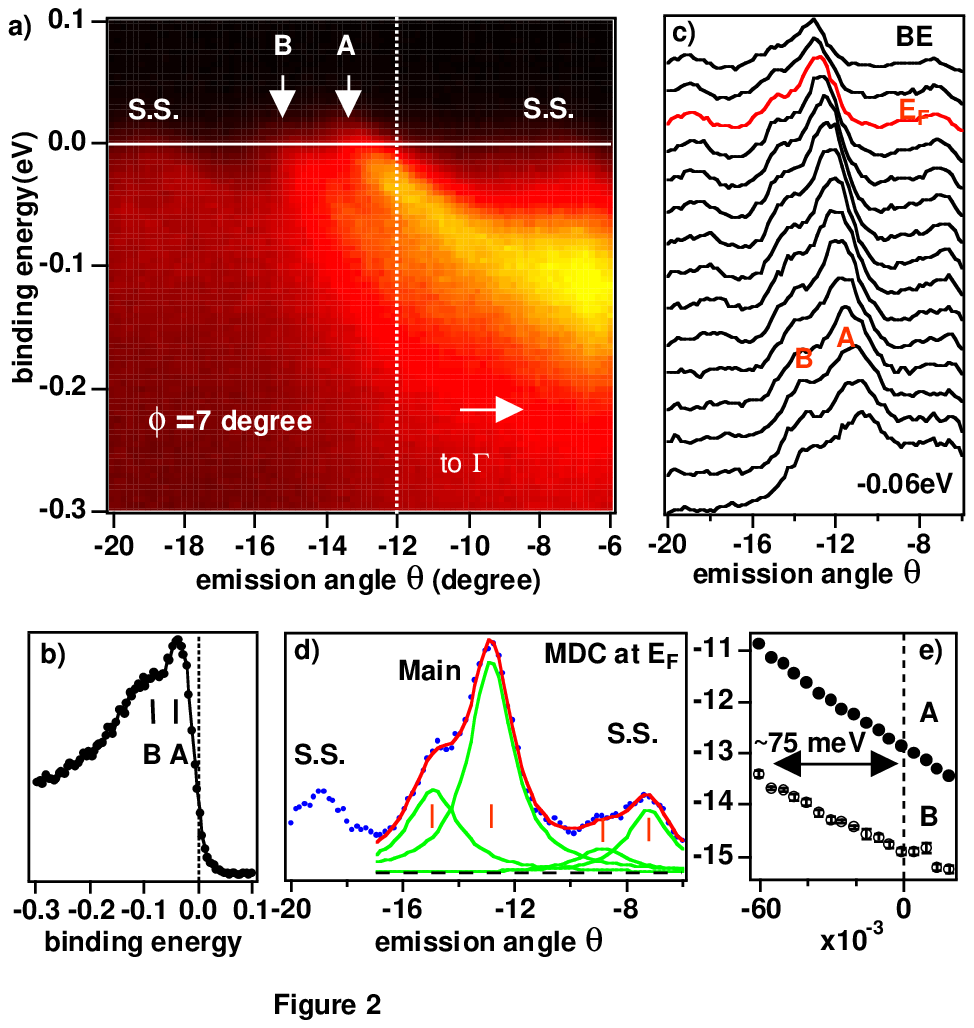}
    \caption{(a) False color plot of $E$ vs $\theta$ for the $\phi=7^{o}$ cut (blue 
  line in figure 1{\bf a}.  (b) EDC at $\theta=-12^{o}$ from panel 
  {\bf a} (vertical 
  white dash line).  Two distinct features, A and B, can be clearly seen in this 
  EDC. (c) MDCs from panel {\bf a} from binding 
  energies $+10 
  meV$ to $-60 meV$, with $5meV$ step. (d) The MDC at $E_{F}$ (blue 
  dots), including a 
  deconvolution of the main band (red line shows the fitting result) into 
  two Lorentzian functions A and B (green lines) plus two
  corresponding features in superstructure band (green lines) on 
  top of a linear background (black dash line). (e) The energy 
  dependence of the $\theta$ value of MDC peaks A (close circles) and B (open 
  circles).  The error bar from the fitting is smaller than the symbol size.  }
 \end{center}
\end{figure}

The energy splitting of these two bands can be most accurately found 
by fitting the MDC peaks for a range of energies.  Figure 2$\bf e$ 
shows the results of the fits from the $\phi=7^{o}$ data of 2$\bf c$.  
The black dots are the peak centroids $(\theta,$ in degree) for each 
of the components of the main band.  The energy splitting can then be 
measured as the separation between the two MDC branches, which is 
immediately seen to be $\sim 75 meV$.

There are several possible origins for the band splitting, both 
intrinsic and extrinsic.  We begin by discussing and ruling out the 
extrinsic effects.  First among these is the possibility that the 
sample surface was slightly multifaceted, which each facet 
contributing to a feature at a slightly different angle.  This is not 
the case, as a post-cleave laser reflection plus optical inspection 
revealed angular deviations no greater than $0.2^{o}$, while the 
angular splitting of the states is on the order of $2^{o}$.  X-ray 
Laue and LEED reflections showed sharp spots as well, indicating high 
quality crystalline of the samples.  We also considered the 
possibility that there may be Bi2201 intergrowths in the Bi2212 
samples, which would be very difficult to detect from magnetization or 
x-ray measurements.  In this possibility, it could be imagined that 
one of the branches was due to Bi2212 and the other to Bi2201.  One 
could possibly imagine this effect causing the splitting on one 
cleave, but not on the numerous cleaves we have studied.  Also, we 
have cooled a sample down to $35K$ which is below $T_{c}$ for Bi2212 
and above the $T_{c}$ for Bi2201.  Since the superconducting gap was 
observed in both branches, it rules out the possibility of the Bi2201 
contamination.

\begin{figure}[htbp]
  \begin{center}
    \includegraphics[scale=0.8]{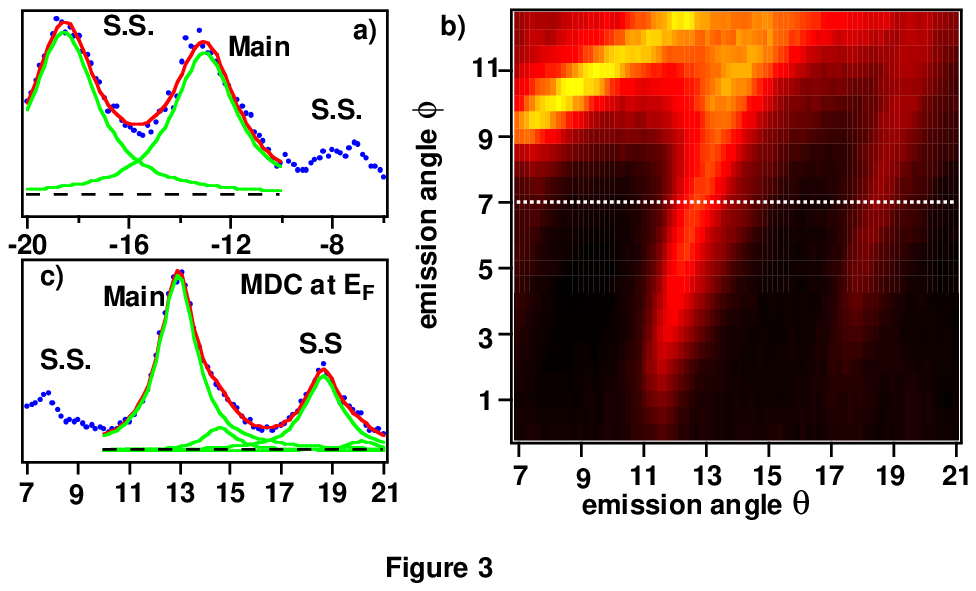}
    \caption{(a) MDC at $E_{F}$ for the $\phi=7^{o}$ cut of lightly overdoped 
	Bi2201, under the identical experimental conditions as figure 2. (b) 
	same data as figure 1{\bf b} except with the polarization $66\%$ along the 
	$\theta$ direction and $34\%$ out of plane.  (c) MDC at $E_{F}$ for the 
	$\phi=7^{o}$ 
	cut (white line in figure 3{\bf b}). }
 \end{center}
\end{figure}

Two main intrinsic possibilities exist: (a) the bilayer splitting due 
to the intracell c-axis coupling, and (b) microscopic phase separation 
into hole rich and hole poor regions each with distinct bands and 
Fermi surfaces.  We first discuss option (a), which we believe to be 
much more likely.  We first remind the reader that in the bilayer 
splitting scenario, the electronic states from each of the two CuO 
planes per unit cell will couple, breaking their degeneracy.  One set 
(antibonding) will increase in energy while the other (bonding) will 
decrease.  Within this picture we make the assignment on the figures 
as peak A=antibonding and peak B=bonding.  Important information is 
garnered by comparing the spectra to that of the single layer compound 
Bi2201, which can not exhibit bilayer splitting.  Figure 3$\bf a$ 
shows an MDC at $E_{F}$ from the $\phi=7^{o}$ angular slice of a 
lightly overdoped Bi2201 sample taken under the identical experimental 
configuration as that of figure 2.  Unlike figure 2$\bf d$, a single 
Lorentzian function alone is enough to produce a very good fit to 
either the main or superstructure band.  This indicates that there is 
no splitting of the bands in Bi2201, at least not one with a similar 
magnitude as observed in Bi2212.  This Bi2201 data also helps rule out 
other possible origins for the band doubling since the crystal 
structures for Bi2212 and Bi2201 are very similar.  For example, if 
the additional band in Bi2212 was a $Bi-O$ band or a 
$Cu_{3d_{3z^{2}-r^{2}}}-O_{2p}$ hybrid \cite{firstprinciple} then we 
would expect to see them showing up in Bi2201 as well.  The lack of 
the apparent doubling in Bi2201 seems to argue against microscopic 
phase separation, however since the mechanism for the phase 
segregation is not completely conclusive, we hold this as a less 
likely mechanism.

The $\vec k$-space dependence of the splitting of the two bands is 
also consistent with the bilayer splitting effect.  Symmetry arguments 
dictate that the coupling between the planes should vanish along the 
$(0,0) \rightarrow (\pi,\pi)$ symmetry line and grow as the $(\pi,0)$ 
region is approached, exactly as the data in figure 1 shows.  
More precisely, the bilayer splitting has been parameterized by 
$\Delta(\vec k) = 0.5*t_{\perp}(cos(k_{x}a)-cos(k_{y}a))^{2}$ which 
has a maximum splitting of $2t_{\perp}$ at $(\pi,0)$ 
\cite{Chakravarty,Anderson}.  Using this equation to parameterize the 
overdoped data of figure 1 and 2, we find a $t_{\perp}$ of $\sim 55 
meV$ and an extrapolated splitting at $(\pi,0)$ of $110 meV$.  This is 
small compared to the band structure prediction of nearly $300 meV$ 
\cite{Massidda} implying important correlation effects present in the 
system, but it is still larger than or comparable to other important 
energy scales $(T_{c}, T^{*}, J)$ meaning that it is an effect which 
should not be ignored for the physics of these compounds.

Although we have not yet carried out a full doping study, our 
preliminary data indicates that near-optimal or underdoped samples 
have a smaller splitting effect.  This smaller splitting, as well as 
the broader peaks associated with less heavy doping, make the 
detection of the splitting much more difficult, probably explaining 
why this effect has not been observed previously.  We also note that 
it would be harder to argue for the existence of the splitting from 
the conventional EDCs because the lineshapes are poorly known.  For 
the MDC's which have only been used extensively in the past year or 
so, the lineshape is simple and additional peaks can be more readily 
seen, even if the peaks are not as sharp or splitting as large as it 
is in the data shown here.  Additionally, much effort has been 
concentrated along either the $(0,0) \rightarrow (\pi,\pi)$ nodal line 
where the splitting vanishes, or near the $(\pi,0)$ point where the 
superstructure bands contaminate the data and give additional peaks.  
Hence for a clear determination of the bilayer splitting effect it is 
important to study the middle of the zone, as done here.

Also relevant to why it may not have been previously detected are the
matrix element effects, which may strengthen one of the peaks at the 
expense of the other.  For example, for the data of figure 1 and 2, a 
large out-of-plane component of the photon polarization was used, 
which helped to enhance the already-weak B peak.  The identical data 
taken with a more in-plane photon polarization shows a much weaker B 
peak, as shown in figure 3$\bf b$ and 3$\bf c$, however the fitting 
result produces same amount of the splitting in energy (data not shown 
here).  The major difference between the data with less grazing photon incidence 
angle (case in figure 3) and more grazing angle (case in figure 1 and 
2) is the relative intensity of peak A and B which makes the 
feature B barely visible in figure 3$\bf b$.  Of course, the relative 
strength of A compared to B can also depend on other parameters, most 
notably the photon energy, for which oscillations in the relative 
magnitude of the branches may exist.

Although the bilayer splitting scenario is convincing, we note that 
there is increasing evidence for microscopic phase separation in BSCCO 
and related compounds, and such an explanation for the band splitting 
is hard to completely rule out \cite{phase-sep}.  However, if this is the explanation a 
couple of important restrictions can be made from this study: (1) The 
degree of phase separation should be larger for Bi2212 than Bi2201.  
(2) There would be metallic regions of two dominant types (possibly 
with other insulating regions) as the MDCs show two main peaks.  (3) 
The angular width of each MDC peak is on the order of $1^{o}$ HWHM or 
$0.04\dot{A}^{-1}$, giving a mean free path for scattering $l=1/\Delta 
k \sim 25\dot{A}$ which sets the minimal domain size in the sample.

From our new data, we see that the B piece of FS matches the 
traditional hole-like piece of FS very well 
\cite{Ding,Borisenko,Fretwell}, while the A piece deviates in $\vec 
k$-space significantly and approaches $\bar{M}(\pi,0)$.  While an 
extrapolation is required, it appears that the A piece probably 
remains centered around $Y(\pi,\pi)$, i.e.  it remains to be 
hole-like.  However, it comes close enough to $(\pi,0)$ that a very 
small amount of $k_{z}$ dispersion of this band could easily push it 
back and forth across $(\pi,0)$, making it either electron-like or 
hole-like.  Experimentally, $k_{z}$ is varied by changing photon 
energies, which also should affect the intensity ratio of the two 
components.  Assuming a small amount of $k_{z}$ dispersion, this would 
give a very natural explanation for the recent controversy over the FS 
topology, in which the existence of an electron-like portion centered 
around $(0,0)$ has been suggested \cite{Chuang,e-DongLai,Pasha} as 
well as denied \cite{Borisenko,Fretwell}.

Such a $k_{z}$ dispersion would come from a c-axis intercell 
coupling, and while this has yet to be directly observed, we note that 
the existence of the intracell coupling shown here makes the 
non-vanishing intercell coupling more plausible.  Recent optical 
measurements on Tl-based cuprates do suggest the presence of this 
interlayer coupling \cite{Tsvetkov}.  Similar to the intralcell 
coupling, symmetry arguments tell us that the intercell coupling 
should vanish along the $(0,0) \rightarrow (\pi,0)$ nodal line and 
should be largest near $(\pi,0)$, precisely where they would need to 
be to resolve the controversy. However, how to connect above argument 
and Bi2201 FS topology at $33eV$ is still not clear to us.

In summary, we have shown that in the normal state of Bi2212 there is 
a doubling of the band structure.  Near the middle of the zone the 
angular splitting is a little more than $2^{o}$ and the energy 
splitting is roughly $75 meV$.  The $\vec k$-dependence of the 
splitting is consistent with that expected from bilayer splitting, and 
if we parameterize this, we obtain a intra-cell coupling $t_{\perp}$ 
of $\sim 55 meV$.  Another less likely origin for the splitting is the 
microscopic phase separation.  Similar results on even more heavily 
overdoped samples have been independently obtained by the Stanford 
group as well \cite{DongLai}.  We acknowledge sample preparation help 
from J. Koralek and beamline support from H. Hochst, S. Kellar and 
D.H. Lu.  This work was supported by the NSF Career-DMR-9985492 and 
the DOE DE-FG03-00ER45809.  SSRL and the ALS are operated by the DOE, 
Office of Basic Energy Sciences and the SRC is supported by the NSF.

\end{multicols}
\vspace*{-0.2in}

\begin{references}
\bibitem{normalization} In this binding energy window, the spectrum 
is featureless in both energy and momentum.
\bibitem{range} Beyond $\phi=8^{o}$ the 
superstructure bands contaminated the data; below $\phi=3^{o}$, the 
splitting is too small such that a single Gaussian or Lorentzian 
function can produce reasonably good fit.
\end{references}
\end{document}